\documentclass[onecolumn,secnumarabic,amssymb,nobibnotes, aps, prd]{revtex4-2}

\usepackage{graphicx}
\usepackage{natbib}
\usepackage{url}
\usepackage{textcase}
\usepackage{xcolor}
\usepackage{bm}
\usepackage[normalem]{ulem}
\usepackage{gensymb}
\usepackage{comment}
\usepackage{lineno}

\def \bii{BiI$_{3}$} 
\def \Fref{Fig.~\ref}

\begin{document}

\title{Unveiling exciton formation: exploring the early stages in time, energy and momentum domain}

\author{Valentina Gosetti$^{1,2,3}$, Jorge Cervantes-Villanueva$^{4}$,
Selene Mor$^{1,2}$, Davide Sangalli$^{5}$, Alberto García-Cristóbal$^{4}$, Alejandro Molina-Sánchez$^{4}$, Vadim F. Agekyan$^6$, Manuel Tuniz$^7$, Denny Puntel$^7$, Wibke Bronsch$^8$, Federico Cilento$^8$, Stefania Pagliara$^{1,2}$}

\email[corresponding author:]{stefania.pagliara@unicatt.it}

\address{$^1$Department of Mathematics and Physics, Università Cattolica, I-25121 Brescia, Italy}
\address{$^2$Interdisciplinary Laboratories for Advanced Materials Physics (I-LAMP), Universit\`a Cattolica, I-25121
Brescia, Italy}
\address{$^3$ Department of Materials Engineering, KU Leuven, Kasteelpark Arenberg 44, 3001 Leuven, Belgium}
\address{$^4$ Institute of Materials Science (ICMUV), University of Valencia,  Catedr\'{a}tico Beltr\'{a}n 2,  E-46980,  Valencia,  Spain}
\address{$^5$ Istituto di Struttura della Materia-CNR (ISM-CNR), Division of Ultrafast Processes in Materials (FLASHit), Area della Ricerca di Roma 1, Monterotondo Scalo, Italy}
\address{$^6$ St. Petersburg State University, St. Petersburg, 199034, Russia}
\address{$^7$ Dipartimento di Fisica, Università degli Studi di Trieste, 34127 Trieste, Italy} 
\address{$^8$ Elettra-Sincrotrone Trieste S.C.p.A., Strada Statale 14, km 163.5, IT-34149 Trieste, Italy} 

\date{\today}%

\begin{abstract}
Resolving the early-stage dynamics of exciton formation following non-resonant photoexcitation in time, energy, and momentum is quite challenging due to their inherently fast timescales and the proximity of the excitonic state to the bottom of the conduction band. 
In this study, by combining time- and angle-resolved photoemission spectroscopy with \emph{ab initio} numerical simulations, we capture the timing of the early-stage exciton dynamics in energy and momentum, starting from the photoexcited population in the conduction band, progressing through the formation of free excitons, and ultimately leading to their trapping in lattice deformations. The chosen material is bismuth tri-iodide (\bii), a layered semiconductor 
with a rich landscape of excitons in the electronic structure
both in bulk and in monolayer form. The obtained results, providing a full characterization of the exciton formation, elucidate the early stages of the physical phenomena  underlying the operation of the ultrafast semiconductor device.
\end{abstract}

\maketitle



\section{Introduction}

Excitons dominate the optical and electronic properties of semiconductors, thus making the fundamental research of how excitons emerge and their following relaxation pathways essential for future developments in optoelectronics, valleytronics and spintronics \cite{Koppens2014, WangG2018, Mueller2018, Boschini2024}.

An exciton is a bound state of an electron excited in the conduction band and a hole left in the valence band. Its formation and relaxation dynamics strongly depend on the photoexcitation condition: when the absorbed photon is on resonance with the excitonic state, it creates an initial coherent population of excitons \cite{Perfetto2016} that evolves towards a decoherent one in a few tens of femtoseconds due to the scattering events with phonons and other degrees of freedom~\cite{Selig2016, Trovatello2020, Mor2021, perfetto2024, Mor2024, Gosetti2024}.
Conversely, when the photon energy exceeds the fundamental bandgap of the semiconductor, a population of quasi-free charge carriers is first excited in the conduction band; only then excitons emerge from this population as a result of the weakly-screened electron-hole Coulomb attraction. The following cascade process that accounts for the exciton relaxation typically involves a plethora of lower-energy, optically bright and/or dark excitonic states whose complex ultrafast dynamics have remained elusive until now \cite{Brem2018, Christiansen2019, Tanimura2019}.

Among the most debated exciton relaxation dynamics, trapping of initially free excitons can take place due to a potential deformation generated by defects in the crystal structure. These excitonic states are referred to as trapped excitons (TE) and are currently object of intense study \cite{Meggiolaro2020, Buizza2021, Dai2024a, Mor2024b}.
In the case of soft semiconductors, excitons can be also self-trapped (STE) by the local elastic lattice distortion generated by strong exciton-phonon coupling.
Initially, the interaction of the excitons with the optical phonons creates large exciton-polarons which are quite mobile; then, due to the interaction with acoustic phonons, excitons become localized on a single unit cell and small exciton-polarons form.
Generally, the trapping process occurs in a few hundred of femtoseconds \cite {Tan2022} and is usually observed in photoluminescence (PL) experiments as a broadband structure below the bandgap \cite{Wu2021, Xu2021, Dai2024a, Dai2024b}.\\
Exciton physics is commonly explored by all-optical spectroscopy methods, which measure direct interband transition, thus allowing to directly access only the bright exciton dynamics.
As a result, the dynamics of optically-dark indirect excitons, the momentum-distribution of the photoexcited charge carriers and excitons, and thus the whole exciton formation and relaxation processes remain inaccessible to the all-optical techniques. 
Time- and angle-resolved photoemission spectroscopy (tr-ARPES) is one of the most successful techniques to resolve nonequilbrium electronic structures in the time, momentum and energy domains.  However, observing a two-particle state such as an exciton is not straightforward, and in the last two decades, only a few tr-ARPES studies have reported excitonic features in either bulk or layered semiconductors like transition metal dichalcogenies \cite{Boschini2024, Weinelt2004, Weinelt2005, Deinert2014, Tanimura2019, Madeo2020, Wallauer2021, Dong2021,  Man2021, Mor2022, Bange2023, Volckaert2023, Tanimura2023}. These works show that the photoemission (PE) signature of an excitonic state reproduces the valence-band momentum dispersion of the holes, as a result of the energy- and momentum-conservation.
Recently, theoretical studies have addressed the ARPES signature of two-fermion quasi-particles such as excitons. Starting from the seminal work by Perfetto \textit{et al.}~\cite{Perfetto2016}, the excitonic signature in tr-ARPES has been modelled both for zero and finite momentum excitons, allowing for a complete characterisation of the signal~\cite{Rustagi2018, Steinhoff2017, Rustagi2019, Christiansen2019, Sangalli2021}. 

In this work, we capture the exciton formation and relaxation dynamics in time, momentum and energy via tr-ARPES in order to shed light on this rich landscape of exciton physics. To achieve this goal, we excite charge carriers in the conduction band of a lead-free layered semiconductor, bismuth triiodide (\bii). In this material, the layered structure gives rise to different series of excitons with Frenkel-like characters and binding energies up to hundreds of meV in the bulk as well as in the monolayer (the lowest bright exciton has a binding energy of 140 meV in bulk and approximately 900 meV in monolayer \cite{Cervantes-Villanueva2024}). Experiments are complemented by numerical simulations, where the tr-ARPES signal of BiI$_{3}$ is reconstructed by combining the prediction of tr-ARPES in presence of excitonic populations, thanks to accurate \emph{ab initio} simulations. We compute a broad excitonic spectrum including states deep into the continuum of the electron-hole Hamiltonian, to capture within the same approach the signal due to both bound excitons and free carriers.

\section{Data analysis and discussion}

\begin{figure}[h]
\includegraphics[width=1\columnwidth]{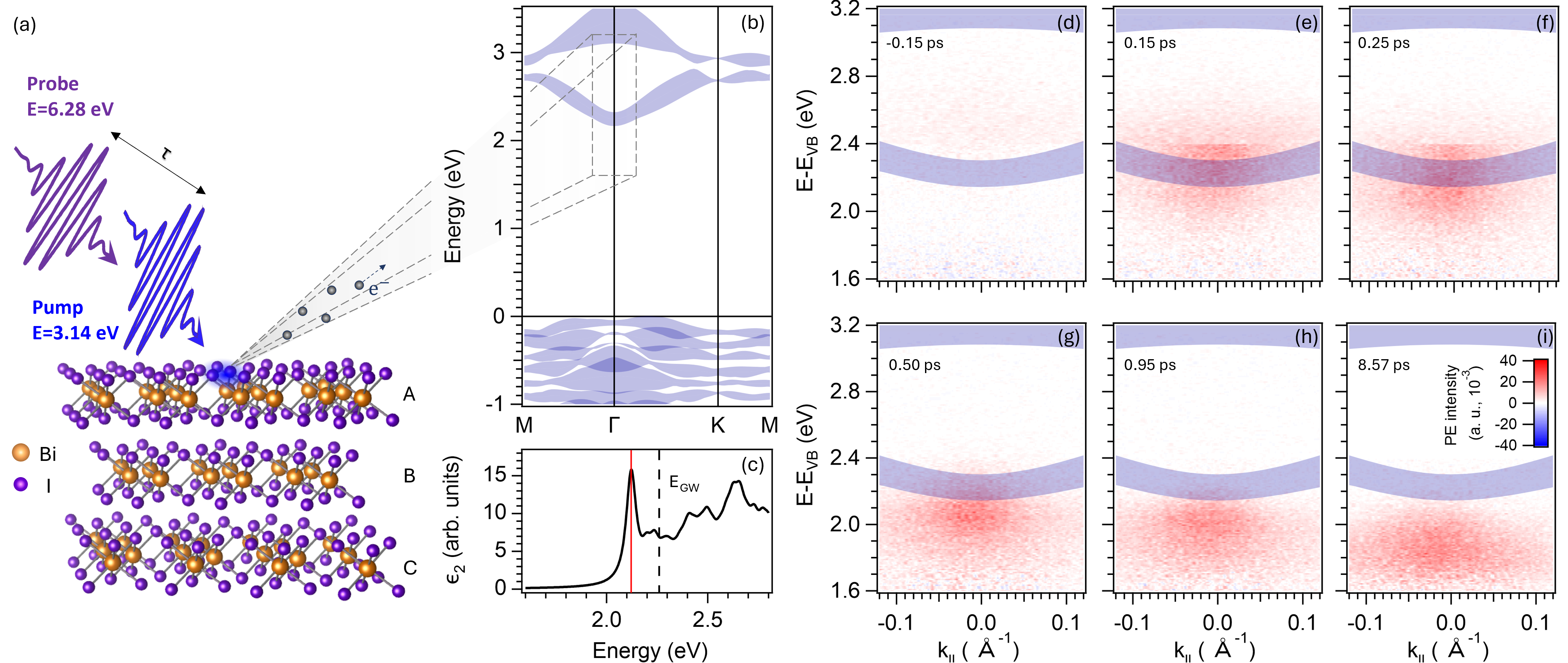}
\caption{(a) Sketch of the tr-ARPES experiment and  \bii\, layered structure. (b) Calculated \bii\, electronic band structure with the integrated dispersion along the $k_{z}$ direction indicated by the shaded areas.  (c) Calculated absorption spectrum, where the red line represents the energetic position of the lowest bright exciton and the dashed line indicates the bandgap calculated with the GW method. (d)-(i) ARPES intensity at selected time delays as a function of intermediate state energy ($E-E_{VB}$) and the parallel momentum, respectively.  The grey-shaded areas represent the computed first and second conduction bands.}
\label{Fig1}
\end{figure}

We study tr-ARPES of a \bii\ single crystal. \bii\,is a layered semiconductor which belongs to the space group $R$-3 and has a double point group symmetry $S_6$.
In the ABC stacking form, the unit cell is rhombohedral, with six iodine atoms and two bismuth atoms (see \Fref{Fig1}(a)). \bii\, presents highly ionic bonds and weak van der Waals interlayer forces. 
The electronic band structure is calculated within the framework of the density functional theory (DFT) \cite{martin2020electronic}, and quasiparticle corrections are applied using the GW method \cite{reining2018gw,onida2002electronic}. The surface-projected band structure is shown in \Fref{Fig1}(b), with surface parallel momentum along a 2D path in the hexagonal Brillouin zone, and shaded areas accounting for the integrated dispersion along the k$_{z}$ direction (see the Supplemental Information (SI) for a detailed explanation). The absorption spectrum (see in \Fref{Fig1}(c)) is computed by solving the Bethe-Salpeter equation (BSE) at zero-momentum, e.g. including excitonic effects, on top of the GW band structure~\cite{strinati1988application}. The energy position of the first bright exciton, 2.12 eV, is marked by a red continuous line.

The tr-ARPES measurements are based on the pump-and-probe scheme, sketched in \Fref{Fig1}(a). The pump photon energy at 3.14 eV is lower than the ionization energy of \bii\,(E$_i$ = 5.8 eV \cite{Tiwari2018}) and much higher than the bottom of the conduction band, in order to enable the excitation of quasi-free charge carriers into the normally unoccupied excited states. 
The s-polarized probe photons at 6.28 eV are used to photoemit electrons from the transiently occupied excited states. This probe photon energy enables spanning in a single measurement the parallel momentum region between -0.12 \AA$^{-1}$\, and 0.12 \AA$^{-1}$ \, and the two lowest-energy conduction bands, as marked by the dashed-line box on top of the calculated band structure in \Fref{Fig1}(b).

\begin{figure}[h]
\includegraphics[width=0.8\columnwidth]{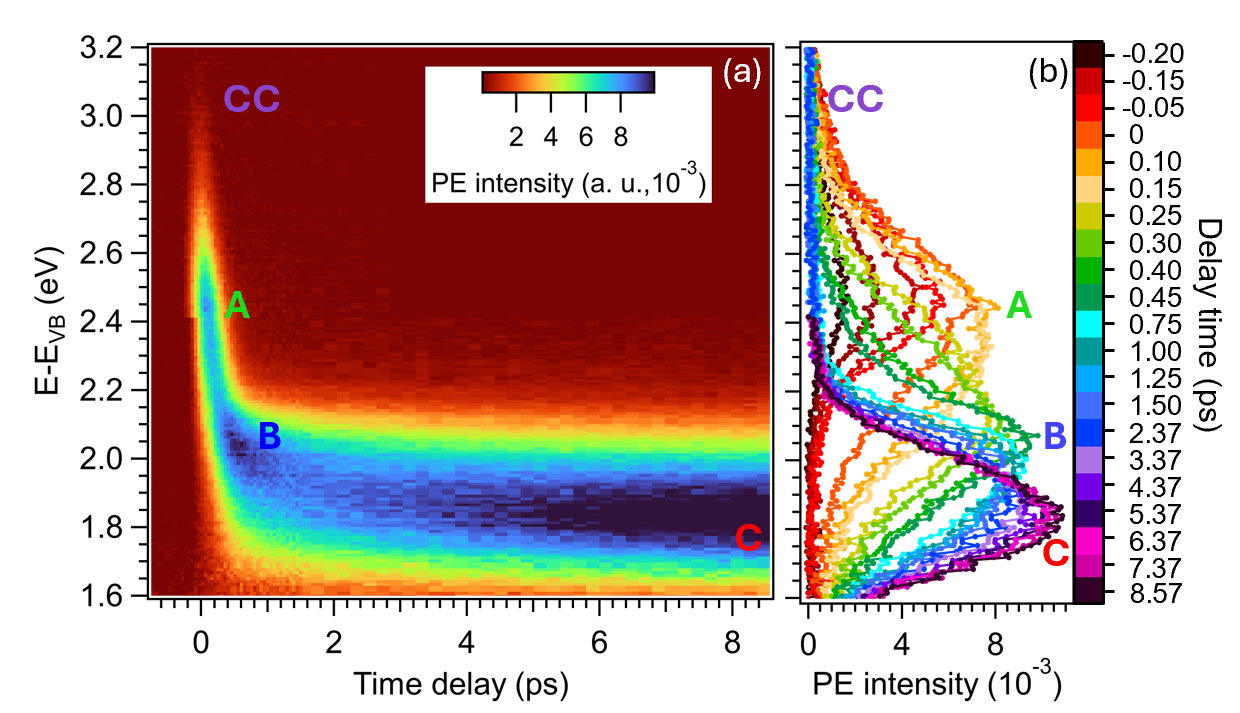}
\caption{(a) PE intensity integrated from the -0.12 \AA$^{-1}$\, to 0.12 \AA$^{-1}$ as a function of intermediate state energy and pump-probe time delay. (b) Vertical linecuts of (a) at selected pump-probe time delays. The resonances are labelled with A, B and C and CC represents the cross-correlation between the pump and the probe.}
\label{Fig}
\end{figure}

The absorbed pump fluence is $6 \frac{\mu \mathrm{J}}{\mathrm{cm}^2}$. Despite the extremely low fluence, we observe a sizable signal. 
\Fref{Fig1}(d-i) shows differential PE intensity maps at several pump-probe time delays after subtraction of the ARPES intensity recorded before the pump pulse arrival.
The data are plotted as a function of parallel momentum ($k_{||}$, x-axis) and intermediate state energy ($E-E_{VB}$, y-axis). 
In order to directly compare the energy of the excited population with the energy of the optical transitions predicted by our calculations, the zero of the intermediate state energy is aligned to the valence band maximum measured by ARPES with a photon energy of 10.8 eV (detailes on the procedure are reported in SI). 

The red color intensity represents a photo-induced increase of electron population with respect to the equilibrium state. 
Before the pump arrival ($t$ = -0.15 ps in \Fref{Fig1}(d)), no features are recorded, which ensures the absence of any residual population with lifetime exceeding the inverse repetition rate of our laser source.   
After the pump pulse arrival ($t$ = 0.15 ps, \Fref{Fig1}(e)), the PE intensity appears as a circular-shaped feature at a kinetic energy of about 2.5 eV and at $k_{||}$=0 \, \AA$^{-1}$. 
To reference the energy position of the photoexcited population in the BiI$_{3}$ electronic band structure, the calculated conduction bands structure is overlaid on top of the data as blue-shades.
We note that the electronic population occupies mainly the lowest-energy conduction band.
At later time delays (\Fref{Fig1} (f-h)), the PE intensity shifts towards lower energies until it sets at 1.8 eV, and eventually it spreads in momentum at 8.57 ps (\Fref{Fig1}(i)). Remarkably, this excited population sits a few hundred meV below the conduction band bottom.

The energy relaxation of the excited electron population is shown in \Fref{Fig}(a), where the integrated PE intensity is plotted as a function of the pump-probe time delay (x-axis) and the intermediate state energy $E-E_{VB}$ (y-axis), after momentum integration between -0.12 \AA$^{-1}$\, and 0.12 \AA$^{-1}$. An initial cross-correlation (CC) PE intensity appears at $\approx$ 3.1 eV, which is resonant with the pump photon energy and the second conduction band. These conduction band electrons relax toward lower energies with two timescales of the order of half-ps and a few ps, respectively. During the relaxation process, the PE intensity shows three resonances, better highlighted by \Fref{Fig}(b) showing a set of energy distribution curves (EDCs) collected at various pump-probe time delays.
The three resonances (A, B, C) appear at approximately 2.5 eV, 2.1 eV, and 1.8 eV.

\begin{figure}[h]
\includegraphics[width=0.6\columnwidth]{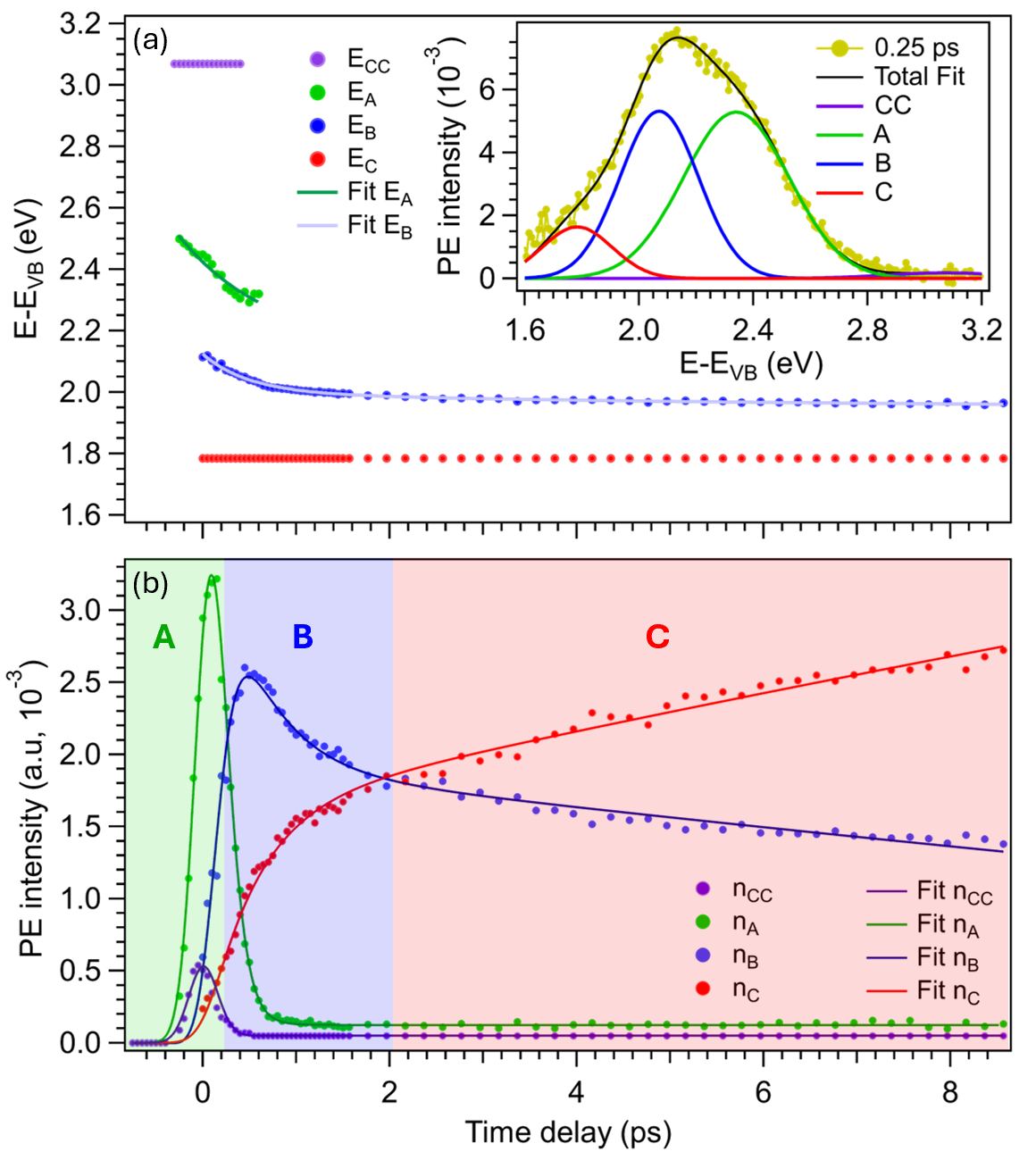}
\caption{Temporal evolution of the energy(a) and the integrated PE intensity(b) of the A, B, and C resonances and the CC feature. The inset shows the EDC at 0.25 ps, the relative four Gaussian fit and the resulting components.}
\label{Fig2}
\end{figure}

We study the temporal evolution of these resonances by fitting the EDCs to a sum of four Gaussian functions, as shown in the inset of \Fref{Fig2} (a). The resulting energy and integrated intensity of each resonance are reported in \Fref{Fig2} (a) and (b) as a function of time. We find that the energy of the A and B resonances shifts towards lower values within the first ps, with the CC and the C energy constant in the fitting procedure. 
All resonances show a delayed intensity rise followed by a multiple-timescale relaxation dynamics that is globally fitted by a convolution of the response function describing the temporal evolution of the PE signal and a Gaussian function, accounting for the experimental resolution.
The analysis of these resonances allows to track the relaxation process of the excited state population.  
After the pump pulse arrival and within the pump-probe CC time, the electronic population relaxes to the first resonance (A) at 2.5 eV corresponding to the bottom of the conduction band. Thus, in agreement with the calculated electronic band structure, the \textit{apparent} energy shift of the A resonance from 2.5 eV to 2.3 eV (green markers in \Fref{Fig2} (a)) is ascribed to the conduction band emptying, which takes place in 120 fs, as also observed in previous time-resolved optical measurements \cite{Brandt2015, Scholz2018, Mor2021}).
 
From the conduction band bottom, the electronic population decays towards the B resonance at $\approx$ 2 eV, thus well within the fundamental bandgap. This is corroborated by the global fitting results which indicate that the intensity build-up time of the B resonance is comparable with the depopulation time of the conduction band $t$ = (120 $\pm$ 30) fs. Then, the electronic population in the B resonance decays and is energy-shifted from 2.1 eV to 1.95 eV on a timescale $t$ = (520 $\pm$ 30) fs. On a longer timescale, exceeding the measured time window, the occupation of the B resonance decays in $t$ = (51 $\pm$ 1) ps at a fixed energy of 1.95 eV. 
The intensity build-up dynamics of the C resonance mirrors the two-steps depopulation of the B resonance and continues to increase until the end of the measurement time window, while its energy is fixed at 1.81 eV.

\begin{figure}[]
\includegraphics[width=1.0\columnwidth]{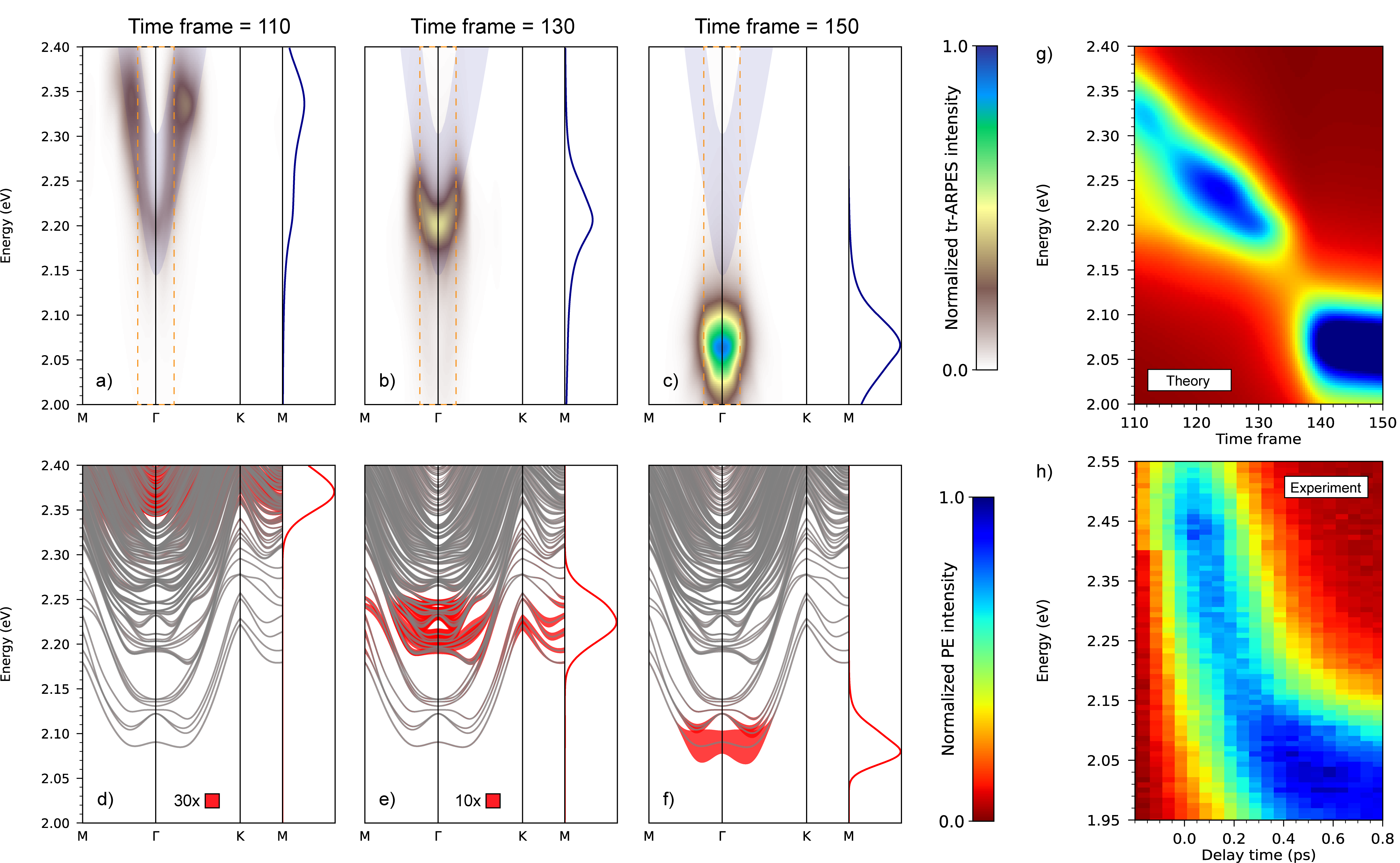}
\caption{(a-c) Theoretical tr-ARPES signal for different time frames. The dashed orange box refers to the area of the reciprocal space measured experimentally. The conduction band minimum with the integrated dispersion along the $K_{z}$ direction is represented by a shaded area. The adjacent panels represent the momentum-integrated PE spectra. (d-f) Exciton dispersion for each time frame, where the occupied excitons are represented in red. The panels on the right show the occupied-exciton density of states. In panels (d) and (e) the contribution is multiplied by the indicated factors for a clearer view. (h-f) Theoretical and experimental PE colormap as a function of the time.}
\label{Fig:trarpes}
\end{figure}

In order to validate the interpretation of the A and B resonance in terms of quasi-free carriers in the conduction band and an excitonic population, respectively, we reconstruct the theoretical tr-ARPES signal starting from an accurate modelling of electronic and optical properties of BiI$_{3}$. First, we compute the exciton dispersion for a broad energy range within the GW@DFT+BSE scheme (see Methods and Ref.~\onlinecite{Cervantes-Villanueva2024} for details). Then, we model the tr-ARPES signal using eq.(4) from Ref.~\cite{Sangalli2021}, which expresses the electron-momentum resolved signal via an integration over the exciton-momentum:
\begin{eqnarray}    
I[N_{\lambda \textbf{q}}(t)](\textbf{k}, \omega) = 2 \pi \sum_{\lambda \textbf{q}} &&N_{\lambda \textbf{q}}(t)
\sum_{cv}  |A^{\lambda \textbf{q}}_{c v \textbf{k}}|^{2} \nonumber \\ 
    &&\times \delta \left(\omega - (\epsilon_{v \textbf{k} - \textbf{q}} + \omega_{\lambda \textbf{q}})\right)
\label{eq:trARPES_theo}
\end{eqnarray}
Similarly to the band structure, the ARPES signal is surface-projected by considering the electron $\mathbf{k}$-parallel momentum. The time dependent population entering the tr-ARPES signal is described in terms of a Gaussian wave-packet initially centered ($t=0$) at the pumping energy, and progressively relaxing towards the exciton dispersion minimum, where it turns into a Boltzmann distribution. Since we consider BSE eigenstates in the continuum of the particle-hole spectrum, both the signal from the conduction band (free carriers) and from the excitonic resonance (bound electron-hole pairs) are captured with a single expression, without the need of introducing independently carriers and exciton populations. The time dependence is determined via few parameters entering the time dependent populations. We do not explicitly fit these parameters to the experimental data, and we report the time dependence in arbitrary units in \Fref{Fig:trarpes}. A detailed explanation of the theoretical implementation is given in the SI.

\Fref{Fig:trarpes} shows the theoretical tr-ARPES signal for different time frames (panels a-c), while the corresponding exciton distribution represented in red on the exciton dispersion (panels d-f). In \Fref{Fig:trarpes}.(a-c), the dashed orange box represents the region of the reciprocal space measured experimentally, and the shaded gray area represents the surface-projected conduction band minimum. Finally, the right-side panels show the $\mathbf{k}$-integrated PE spectra. In \Fref{Fig:trarpes}(d-f) the right-side panels show the $\mathbf{q}$-integrated exciton density of states. The theoretical and experimental 2D PE maps as a function of the time are also shown in panels (g-h). 

The tr-ARPES signal shown in \Fref{Fig:trarpes}.(a-b) is generated by BSE eigenstates within the energy range of 2.4 eV to 2.2 eV, as depicted in Fig. 4(d-e).
Despite being generated by excitons spread over all the Brillouin zone, and without any constrain on the momentum of the photo-emitted electron, the signal is mostly centered around the $\Gamma$ point, e.g. in the window which is detected experimentally. Moreover, at early times, the signal precisely overlaps with the conduction band minimum, indicating that it is actually due to the free carriers located at the conduction band. This signal corresponds to the A resonance in Fig 3(a), which is associated with the relaxation dynamics of the free carriers along the conduction band. Over time, the signal continues to evolve, leading to the signal represented in \Fref{Fig:trarpes}.(c), created by the BSE eigen-states below 2.2 eV, as shown in \Fref{Fig:trarpes}.(f). Since this signal is located below the conduction band, it consists solely of bound excitons. In this case, the signal corresponds to the B resonance shown in Fig 3(b), where the exciton cascade captured experimentally is attributed to the contribution of the lower energy excitons.

Notably, there is an energy region with no intensity between the signals of \Fref{Fig:trarpes}.(b) and \Fref{Fig:trarpes}.(c). This is easier to appreciate when comparing the calculated and experimental tr-PE colormap in \Fref{Fig:trarpes}.(g) and (h), respectively. The absence of PE signal is due to the lack of excitonic states in the energy range around 2.15 eV, as shown in the exciton dispersion reported in Fig 4(d-f). This gap corresponds to the energy jump between the A and B resonances in Fig 3(a), although the theoretical energy jump is smaller than the experimental one. We attribute this discrepancy to surface effects which are neglected in the simulations. Indeed, the signal is collected from the first few surface layers where a reduced screening is experienced, due to the presence of the vacuum. Besides this quantitative difference, the results of the simulations strongly corroborate the identification of the A and B resonances.

\begin{figure}[h]
\includegraphics[width=0.5\columnwidth]{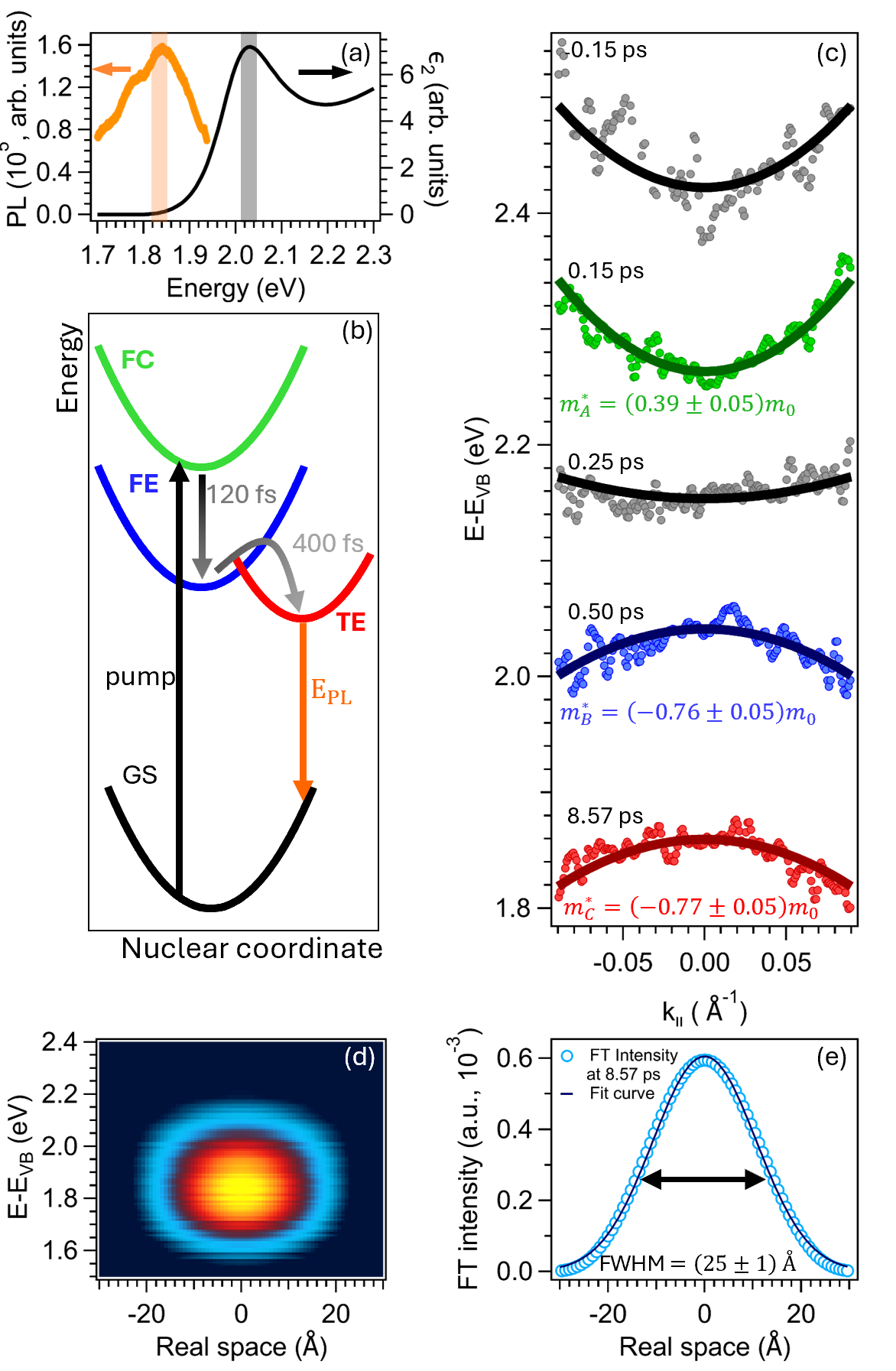}
\caption{(a) PL measurement at RT (orange) and calculated imaginary part of the dielectric function taken from \cite{Mor2021} (black). The respective peak positions are marked by vertical dashed lines.(b) Diagram of the exciton formation and trapping process (c) Intermediate energy vs parallel momentum dispersion for each frame of \Fref{Fig1}(d-i). (d) FT of the ARPES map collected at 8.57 ps (e) Horizontal linecut of (d) at the maximum of the intensity.}
\label{Fig5}
\end{figure}
Having established the nature of the A and B resonances, we are left with the origin of C resonance. Two aspects characterize this feature: 
(i) its energy is much lower than the lowest-energy free exciton (FE) calculated at $\Gamma$ for the layered rhombohedral \bii\, crystal lattice; (ii) the maximum intensity is reached 520 fs after the pump arrival and delayed by 400 fs with respect to the FE formation (maximum of the B resonance). 
These observations suggest that the C resonance could result from a trapping process of initially free excitons.
\cite{Kaifu1988, Kawai1989, Lifshitz1995, Nila2017, Yasunami2021, Agekyan2022}. To discuss the occurrence of this trapping dynamics, we first note that, depending on the excitation conditions and the sample quality, different kinds of trapped excitons (TE) could appear in the PL spectrum of \bii\, which is shown in \Fref{Fig5} (a) (orange line). TE usually lead to broadband PL emission characterized by a huge Stoke shift (above 100 meV) with respect to the FE \cite{Xu2021, Wu2021, Dai2024a, Dai2024b}. Indeed, our PL measurement shows a broad PL spectrum with Stokes shift of 150 meV, as estimated from the energy separation between the PL and FE absorption peak. Notably, the Stokes shift is comparable with the energy difference between the B and C resonance (0.18 eV) (details on the PL experimental setup are reported in SI). 
Thus, while we can exclude that excitons can be trapped in stacking fault centres as they would have appeared only at very low temperatures (5K) as sub-meV narrow lines very close (10-20 meV) to the FE peak \cite{Kawai1989, Agekyan2022}, excitons can be trapped in other native defects \cite{Lifshitz1995, Nila2017, Agekyan2022} or self-trapped in the local photoinduced lattice distortion. 
In fact, in soft materials like layered semiconductors, the exciton-phonon coupling is strong enough to produce a lattice deformations with a consequent localization of FE as STE \cite{Wu2021, Kastl2022}.
In this regard, \bii\, results to be an ideal platform to host STE, since it is a layered material where FE have Frenkel-like character \cite{Kaifu1988}, large binding energy, and strong coupling with an $A_{1g}$ optical phonon mode  \cite{Scholz2018, Mor2021, Mor2024}. 

We then summarize the exciton formation and relaxation dynamics in the sketch of \Fref{Fig5} (b). 
Initially, the photoexcited quasi-free charge carriers in the conduction band bind to holes and form a population of excitons on a 120-fs timescale. Then, FE undergoes a fast trapping process within 400 fs, likely with the aid of coupled A$_{1g}$ optical phonons, or due to native defects which act as trapping sites.

Finally, we extend our study to the momentum domain by following the evolution of the $k_{||}$-dispersion of the photoexcited population during the relaxation dynamics. For each frame shown in \Fref{Fig1}(d-i), we evaluate a set of EDCs at various $k_{||}$ between -0.09 \AA$^{-1}$\, and 0.09 \AA$^{-1}$\,  through fitting with a Gaussian function. 
The Gaussian peak energy is then shown as a function of $k_{||}$ in \Fref{Fig5}(c) (color markers) in order to appreciate, at each time delay, the momentum dispersion of the photoexcited population.
During the pump-probe CC (i.e. at $t$ = -0.15 ps, grey markers), the conduction band electron population distribute over a parabolic band with upward dispersion. This behavior is observed until t=0.15 ps (green markers), when the population reaches the bottom of the conduction band (A resonance), with an effective mass of $m^*_A=(0.39 \pm 0.05)\,m_0$, where $m_0$ is the electronic mass. The theoretical effective mass of the surface projected conduction band is $m^*=0.58\,m_0$ (k$_{z}$ resolved effective masses range between $0.424\, m_{0}$ and $1.046\, m_{0}$, see SI). At $t$ = 0.25 ps, the $k_{||}$-dispersion flattens significantly. By comparison with \Fref{Fig2}(b), we note that at this time delay the PE intensity of the A resonance equals that of the B-resonance. Therefore, the nearly flat dispersion must be connected to the evolution from a bare conduction-band population to the free-exciton population for which a momentum-dispersion is expected. Indeed, at $t$ = 0.5 ps, i.e. when the PE intensity from the B resonance is maximum, the $k_{||}$-distribution has evolved into a downwards-dispersing parabola with effective mass of $m^*_B=(-0.76  \pm 0.05) m_0$ (blue markers in \Fref{Fig5}(c)). This further corroborates the excitonic nature of the B resonance.
For direct gap semi-conductors, according to eq.~(\ref{eq:trARPES_theo}), the PE excitonic signature evolves from the upward excitonic dispersion (high temperature exciton population) to the downward hole-like parabolic dispersion of the valence band around $\mathbf{k}=\Gamma$ (low temperature exciton population) \cite{Perfetto2016,Sangalli2021,Dong2021}. For an indirect gap material at low temperature the dispersion of the valence band maximum around $\mathbf{k}=\mathbf{q}_{min}$, corresponding to the lowest energy exciton, is expected~\cite{Rustagi2018}. The case of \bii\, is more complex, due to the presence of many electronic and excitonic bands very close in energy  (see SI). The analysis of the theoretical B resonance shows that the effective mass measured experimentally cannot be directly associated with a specific excitonic or valence band, since many contributions are involved in the formation of the signal. However, a concavity switch is observed between A and B resonances in \Fref{Fig:trarpes}(b) and \Fref{Fig:trarpes}(c), respectively, in agreement with the experimental results (see more details in SI).
At even later times  ($t$ = 8.57 ps, red markers), the exciton trapping process maintains the downwards-dispersing parabolic distribution with a comparable effective mass of $m^*_C=(-0.77  \pm 0.05) m_0$.

At this delay time, we obtain the \bii\, exciton distribution in real space, which provides the exciton wave function, through Fourier Transformation  (FT) of the momentum-resolved PE intensity. The procedure is taken from \cite{Dong2021} (see SI for further details). The FT intensity is shown in the 2D-plot of \Fref{Fig5}(d) as a function of real space (x-axis) and intermediate state energy (y-axis). We point out that the intensity at this time delay captures the contribution of co-existing FE and TEs. 
By taking a horizontal linecut at the FT intensity maximum, a Gaussian lineshape width and a FWHM of (25$\pm$ 1) \AA\,is obtained (\Fref{Fig5} (e)), which is comparable with the literature \cite{Dong2021, Volckaert2023}. Considering the volume of a \bii\, octahedral cluster consisting of one bismuth atom and six iodine atoms, we estimate that the exciton wave function spreads over ca. seven clusters.

\section{Conclusion}
In conclusion, we have accurately reconstructed the early-stage exciton dynamics of the layered semiconductor \bii\, in the multidimensional space by a joint experimental and theoretical effort. 
We have found that an initial population of quasi-free charge carriers excited in the conduction bands evolves within 120 fs into an in-gap population of free-excitons at $k_{||} = 0$. These findings have been validated by 
the theoretical reconstruction of the tr-ARPES signal using a GW@DFT+BSE scheme as a starting point. Concomitantly, the momentum dispersion changes from upward (electron-like), when the electron population is in the conduction band, to downward (hole-like), corresponding to the exciton formation and its consequent localization into trapped excitons.   
From the $k_{||}$-dispersion, we calculate the exciton wave function in real space, which 
spread over ca. seven \bii\, clusters.

We emphasize that the possibility to achieve a comprehensive characterization of the early-stage exciton dynamics triggered by light is the first step towards the optical control of ultrafast physical processes in matter.  Harnessing such processes in future device engineering will enable the transmission and manipulation of information  with unprecedented ultrafast switching times.

\section{Methods}
\subsection*{Crystal growth}
The single crystal of \bii\ was grown following the subsequent two-step procedure.
Firstly, a crystalline \bii\ powder is synthesized from a mixture of chemically pure bismuth and iodine powders (weight ratio 2:1). The synthesis is carried out at a temperature of about 200°C in a glass ampoule, the vacuum is $10^{-4}$ mm Hg.
Secondly, the crystalline \bii\ powder is placed in an evacuated ampoule with a vacuum of $10^{-4}$ mm Hg. The ampoule is inserted into the furnace, where the powder is slowly (within an hour) heated up to 300°C. Then, a temperature gradient is created along the ampoule, and \bii\ crystal plates grow in the cold part of the ampoule for one hour. The crystal axis is oriented perpendicular to the plane of the plates.

\subsection*{Sample characterization}

The photoluminescence (PL) measurement is obtained with a Renishaw inVia microscope upon illumination with a continuous-wave laser beam at 633 nm focused down to 1 $\mu$m spot diameter. The beam intensity is reduced to $6 \cdot 10^3 \frac{Watt}{cm^2}$ to prevent sample damage. 
To block the Rayleigh scattering intensity, the microscope is equipped with an edge filter. A diffraction grid enables the detection of Raman shifts above 100 cm$^{-1}$ with 0.9 cm$^{-1}$ spectral resolution. 
To evaluate the position of the first bright exciton we collected the equilibrium reflectance exploiting a PerkinElmer Lambda 950 spectrometer at room temperature. The imaginary part of the dielectric function have been extracted fitting the equilibrium reflactance with a Tauc-Lorentz procedure (detailed explanation in the SI).

\subsection*{Photoemission measurements}

The time-resolved angle-resolved photoemission spectroscopy (tr-ARPES) experiments were carried out in the T-ReX laboratory (FERMI, Trieste).
To explore the non-occupied states in \bii,  tr-ARPES measurements have been performed using a pump photon energy of 3.14 eV and a probe photon energy of 6.28 eV, near the ionization energy of \bii.
The tr-ARPES setup exploits a femtosecond pulsed RegA laser system with a 250 kHz repetition rate (RR), 50 fs pulse duration, and 800 nm central emission wavelength. The RegA output is divided by a beam splitter into two beams. 
The first branch, used a pump, is doubled before entering the photoemission chamber. Through a composite half-wave plate and polarizer, the polarization and the beam intensity can be tuned.
The second part of the beam is used as a probe. We exploited the fourth harmonic generation to have a final photon energy of 6.28 eV. The beam is compressed with a pair of ultra-violet-fused silica (UVFS) prisms. A delay stage is inserted in the pump path to manipulate the pump-probe time delays. 
The beams enter the UHV chamber from a window at 35° with respect to the normal of the sample. The sample faces the analyzer slit and the photoemitted electrons are acquired with a SPECS Phoibos 225 hemispherical analyzer.
The instrumental response function, defined as the cross-correlation between the pump and the probe measured at the sample position, is limited by the temporal duration of the pump pulse and amounts to 200 fs.
To evaluate the energy resolution we have fitted the edge of a polycrystalline gold collected at room temperature. The edge has a Fermi-Dirac shape. We fitted it with a convolution with the Gaussian response function that represents the instrumental response. The resulting energy resolution is 50 meV.
In our experiment, the absorbed fluence is $6\frac{\mu J}{cm^2}$, since the spot size is 150 $\mu m$. The pump and the probe beams are linearly polarized
The sample is kept at 130 K and cleaved \textit{in\,situ}.\\

\subsection*{Theory}

\emph{Ab initio} calculations of the electronic structure of single crystal BiI$_{3}$ were carried out at the DFT level, as implemented in the Quantum ESPRESSO package \cite{giannozzi2009quantum,giannozzi2017advanced,giannozzi2020quantum}. For the exchange-correlation functional, the Perdew-Burke-Ernzerhof (PBE) \cite{perdew1996generalized} was used. The van der Waals interactions are accounted employing the Tkatchencko-Scheffler model (TS-vdW) \cite{PhysRevLett.102.073005}. Fully relativistic norm-conserving pseudopotentials were used to include spin-orbit effects \cite{van2018pseudodojo}. In addition to the corresponding valence states, the $4d$ semi-core states for iodine and the $5d$ for bismuth are included in the pseudopotentials given their importance when calculating the quasiparticle correction \cite{Cervantes-Villanueva2024}. Experimental lattice parameters were used, with lattice constants: $a = 7.52$ \r{A} and $c = 20.71$ \r{A} \cite{nason1995growth} . Quasiparticle corrections in a single-shot GW approach (G$_{0}$W$_{0}$) \cite{stan2009levels} and optical and excitonic properties solving the Bethe-Salpeter equation were obtained using the Yambo code \cite{marini_yambo_2009,sangalli2019many, marsili2021spinorial}. For the GW calculations, the dielectric cutoff was set to 7 Ry, where the dynamical screening effect was treated using the plasmon-pole approximation \cite{PPA_yambo}. For the study of the optical properties including excitonic effects, the BSE is solved on top of the G$_{0}$W$_{0}$ results within the Tamm-Dancoff approximation \cite{strinati1988application}. Non-vanishing momentum transfer excitons were computed solving the finite momentum BSE, where nine valence bands and four conduction bands were considered to construct the dielectric function. A k-grid of 8x8x8 was used in all the calculations, where a q-grid of 8x8x8 was used for the finite momentum BSE, following the Monkhorst-Pack method \cite{monkhorst1976special}. See Ref \cite{Cervantes-Villanueva2024} for a more detailed explanation of the computational parameters. To theoretically reconstruct the tr-ARPES signal, we use the eq.(4) from Ref \cite{Sangalli2021}, where the momentum dispersion of the band structure is projected into the surface to compare with the experimental measurements. In order to define the dynamics of the excitonic population, we define a time-dependent distribution function that is at the beginning dominated by a Gaussian wave-packet initially centered (t = 0) at the pumping energy, which evolves into a Boltzmann distribution as the system progressively relaxes toward the minimum of the exciton dispersion. See the SI for a complete explanation of the theoretical implementation.

\section{Acknowledgements}
S.M. and S.P. acknowledge partial support from D.1 and D2.2 grant of the Universitá Cattolica del Sacro Cuore.
D.S. and A.M.-S. acknowledge funding from the TIMES project thorugh the European Union’s Horizon Europe research and innovation programme under the Marie Sklodowska-Curie (Grant Agreement 101118915).
D.S. acknowledges funding from MaX "MAterials design at the eXascale” co-funded by the European High Performance Computing joint Undertaking (JU) and participating countries (Grant Agreement No. 101093374)
A.M.-S. acknowledges funding from the project of I+D+i PID2020-112507GB-I00 QUANTA-2DMAT and PID2023-146181OB-I00 UTOPIA, funded by MCIU/AEI/10.13039/501100011033/FEDER, project PROMETEO/2021/082 (ENIGMA) and SEJIGENT/2021/034 (2D-MAGNONICS), funded by the Generalitat Valenciana, the Advanced Materials programme (project SPINO2D), supported by MCIN with funding from European Union NextGenerationEU (PRTR-C17.I1) and by Generalitat Valenciana. A. M.-S. acknowledges the Ramón y Cajal programme (grant RYC2018-024024-I; MINECO, Spain). J. C.-V. acknowledges the Contrato Predoctoral Ref. PRE2021-097581.

\section{Author contribution}
SP and DS conceived the study and designed the experiments. VG, DP, MT, WB and FC carried out the experiments.
VG and SM analyzed the data and JCV and AMS calculated the ARPES signal by GW@DFT+BSE scheme. JCV and VG wrote the manuscript and provided equally to the study. VFA supplied the samples. 
All authors contributed to the interpretation of the data, discussed the results and commented on the paper.

\section{Competing interests}
The authors declare no competing interests.

\end{document}


\title{Supplemental Information\\
Unveiling exciton formation: exploring the early stages in time, energy and momentum domain
}



\author{Valentina Gosetti$^{1,2,3}$, Jorge Cervantes-Villanueva$^{4}$,
Selene Mor$^{1,2}$, Davide Sangalli$^{5}$, Alberto García-Cristóbal$^{4}$, Alejandro Molina-Sánchez$^{4}$,
V. F. Agekyan$^6$, Manuel Tuniz$^7$, Denny Puntel$^7$, Wibke Bronsch$^8$, Federico Cilento$^8$, Stefania Pagliara$^{1,2}$}


\address{$^1$Department of Mathematics and Physics, Università Cattolica, I-25121 Brescia, Italy}
\address{$^2$Interdisciplinary Laboratories for Advanced Materials Physics (I-LAMP), Universit\`a Cattolica, I-25121
Brescia, Italy}
\address{$^3$ Department of Materials Engineering, KU Leuven, Kasteelpark Arenberg 44, 3001 Leuven, Belgium}
\address{$^4$ Institute of Materials Science (ICMUV), University of Valencia,  Catedr\'{a}tico Beltr\'{a}n 2,  E-46980,  Valencia,  Spain}
\address{$^5$ Istituto di Struttura della Materia-CNR (ISM-CNR), Division of Ultrafast Processes in Materials (FLASHit), Area della Ricerca di Roma 1, Monterotondo Scalo, Italy}
\address{$^6$ St. Petersburg State University, St. Petersburg, 199034, Russia}
\address{$^7$ Dipartimento di Fisica, Università degli Studi di Trieste, 34127 Trieste, Italy} 
\address{$^8$ Elettra─Sincrotrone Trieste S.C.p.A., Strada Statale 14, km 163.5, IT-34149 Trieste, Italy}  

\date{\today}%

\maketitle
\section*{Experimental results}

\subsection*{Sample Characterization}

\begin{figure}
\includegraphics[width=0.5\columnwidth]{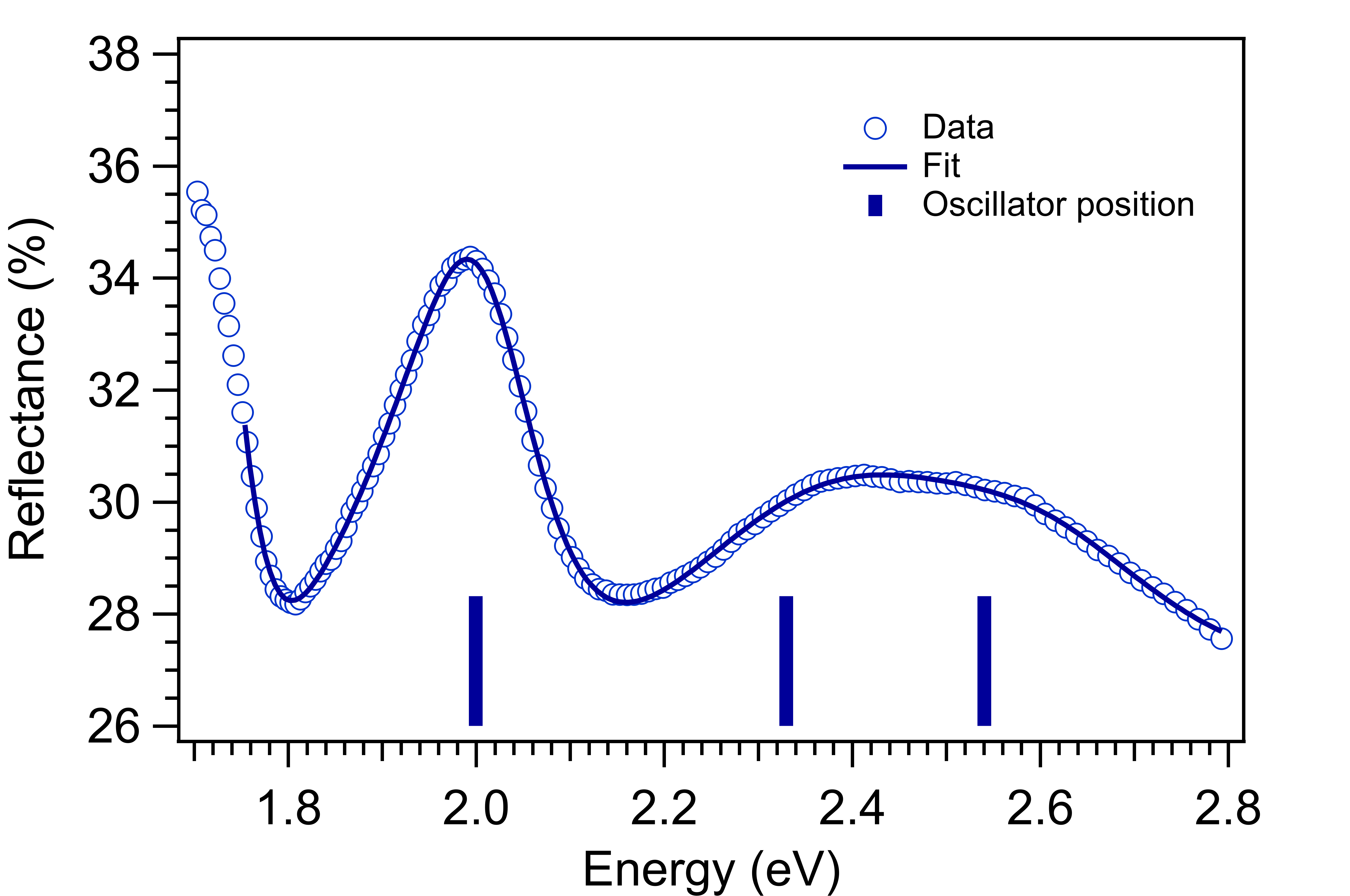}
\caption{Equilibrium reflectance at room temperature with the Tauc-Lorentz fit. The markers represent the position of the oscillators.}
\label{FigS1}
\end{figure}

To evaluate the position of the first bright exciton, we collected the equilibrium reflectance at room temperature exploiting a PerkinElmer Lambda 950 spectrometer. Data are shown in \Fref{FigS1}.
The spectrum exhibits a sharp resonance at 2 eV \cite{Kaifu1988}, ascribed to the lowest-energy exciton, and a broader feature centred at 2.48 eV \cite{Jellison1999}.
The non-zero signal below 1.8 eV is due to the backscattering from the substrate, as discussed in previous works \cite{Mor2021, Mor2024}.
To calculate the imaginary part of the dielectric function ($\epsilon_i$) and find the peak positions, a Tauc-Lorentz model has been chosen to reproduce $R_{eq}$.
This model uses the Tauc joint density of states, which represents the inter-band transitions in a direct gap material, and Lorentzian oscillators to model resonances in the absorption spectrum \cite{jellison1996}. 
In this model, the complex dielectric function is described by:
\begin{equation}
    \tilde{\epsilon}_{TL}=\epsilon_{r,TL}+i\,\epsilon_{i,TL} = \epsilon_{r,TL}+i\,\epsilon_{i,T}\,\epsilon_{i,L}.
\end{equation}
The imaginary Tauc-Lorentz term ($\epsilon_{i,TL}$) is the product of the imaginary Tauc ($\epsilon_{i,T}$) and the imaginary Lorentian ($\epsilon_{i,L}$) dielectric functions.  
In our case, three Tauc-Lorentz oscillators have been exploited to model the excitonic resonances in \bii.
The complete imaginary part of the dielectric function is:

\begin{equation}
\epsilon_{i,TL} =  \epsilon_{i,T} \epsilon_{i,L} = 
\begin{cases}
\sum_{i=1}^{3} \dfrac{1}{E} \dfrac{A_iE_{0i}\Gamma_i (e-E_g)^2}{(E^2-E_{0i}^2)^2+\Gamma_i^2E^2} & \text{if $E>E_g$}\\ 
0 & \text{if $E\leq E_g$},
\end{cases} 
\end{equation}

where $A_i$, $\Gamma_i$, and $E_{0i}$ are the amplitude, the full-width-half-maximum (FWHM), and the centre of the \textit{i}-th oscillator, respectively.
The real part of the dielectric function is retrieved by the Kramer-Kronig transformation: 

\begin{equation}
    \epsilon_{r,TL}(E)=\epsilon_r(\infty)+\frac{2}{\pi}P\int_{E_g}^{\infty} \frac{\xi \epsilon_{TL,i}(\xi)}{\xi^2-E^2}d\xi,
\end{equation}

where P is the Cauchy principal value.
The reflectance is then defined as:
\begin{equation}
    R=\frac{(n-1)^2+k^2}{(n+1)^2+k^2},
\end{equation}

where n and k are the real and imaginary parts of the complex refractive index $\tilde{n}=n+ik$.
These are related to the dielectric function as:
\begin{equation}
 n=\{\dfrac{1}{2}[(\epsilon_1^2+\epsilon_2^2)^{\frac{1}{2}}+\epsilon_1]\}^{\frac{1}{2}} 
\end{equation}
 
\begin{equation}
  k=\{\dfrac{1}{2}[(\epsilon_1^2+\epsilon_2^2)^{\frac{1}{2}}-\epsilon_1]\}^{\frac{1}{2}}.
\end{equation}
Thereby, the reflectance at room temperature has been fitted using the Tauc-Lorentz procedure. 
The resulting fit is shown as a dark blue solid curve in \Fref{FigS1}, where the bars refer to the center positions of the main Tauc-Lorentz oscillators. 
The imaginary part of the dielectric function is the sum of the four Tauc-Lorentz structures and it is then reported in Fig.5 in the main text.

\subsection*{Time resolved photoemission measurements}
The angle-resolved photoemission spectroscopy (ARPES) experiments were carried out at the T-ReX laboratory (FERMI, Trieste).

To determine the edge of the \bii\, valence band position,  a photon energy of 10.8 eV was used. This is well above the ionization energy of \bii\, (5.8 eV \cite{Tiwari2018}). The generation of 10.8 eV photon energy starts with the output of a Coherent Monaco laser system \cite{Peli2020}. This is a Yb-based fiber laser and emits pulses with a central wavelength of 1035 nm, a pulse duration of 290 fs, and a repetition rate (RR) up to 1 MHz. The beam of the probe path is frequency-doubled. Then, the fundamental and the doubled beams interact in a BBO crystal to generate the third harmonic of the fundamental light, by sum frequency generation. The beam is then focused in a chamber filled with Xenon gas where, through a four-wave mixing third harmonic generation (THG) process, the ninth harmonic at 10.8 eV is generated. The ultra-violet (UV) light hits the sample at 30° with respect to the normal emission. The sample faces the analyzer slit and the photoemitted electrons are acquired with a SPECS Phoibos 225 hemispherical analyzer. The sample is mounted on a six-degrees-of-freedom cryo-manipulator. The overall energy resolution is 26 meV. 
The parallel momentum is calculated using the conservation laws for the photoemission process \cite{Damascelli2004, Sobota2021, Zhang2022, Boschini2024}.


\begin{figure}
\includegraphics[width=0.7\columnwidth]{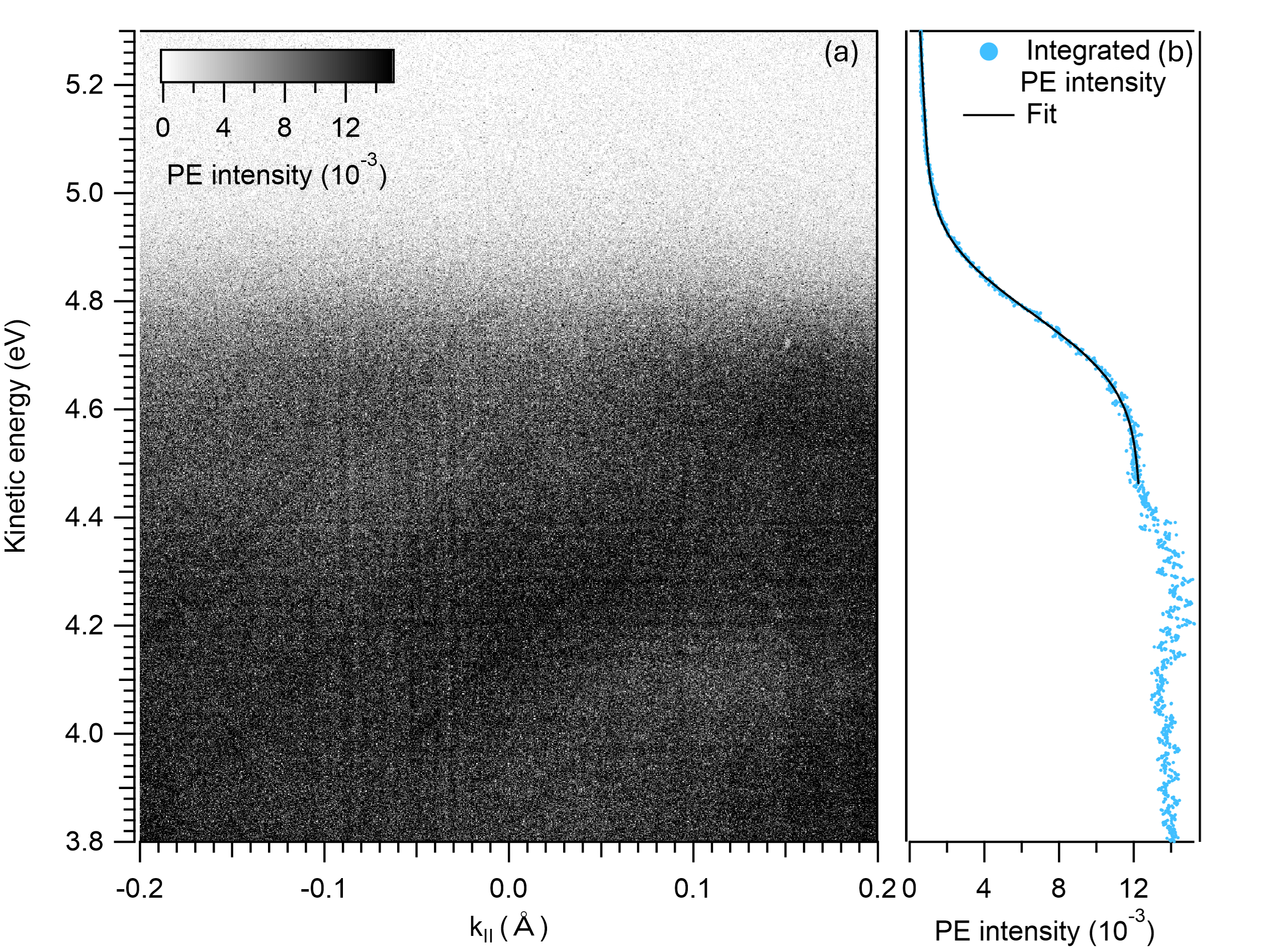}
\caption{(a) ARPES measurement with 10.8 eV photon energy. (b) Integrated PE between -0.2 \AA$^{-1}$ and 0.2 \AA$^{-1}$ as a function of KE energy. }
\label{FigS4}
\end{figure}

To evaluate the energy position of the edge of the valence band, ARPES measurements with 10.8 eV were performed.
The resulting map is shown in \Fref{FigS4} (a), where the PE intensity is plotted as a function of kinetic energy ($E_K$) and parallel momentum (k$_\parallel$). The photoemission intensity shows an increase around $E_K$=4.8 eV due to the valence band.
We integrated the intensity from -0.2 \AA$^{-1}$ to 0.2 \AA$^{-1}$ as shown in \Fref{FigS4}(b), and the obtained curve was fitted by an error function. The central value of the error function, $E_K$=4.77 eV defines the valence band edge. 
Time-resolved experiments were performed using a probe photon energy of 6.2 eV, obtained as the fourth harmonics of a Ti:Sapphire regenerative amplifier (Coherent RegA). 
To set the valence band edge in the photoemission data collected at 6.2 eV (Fig. 1 (d-i) in the main text), we have downshifted the kinetic energy by 4.6 eV, which is the difference between the two photon energies (6.2 eV and 10.8 eV).
Moreover, to compare the photoemission data reported in Fig. 1 in the main text with the optical one, we chose to refer the energy to the valence band edge $E_V$ and to introduce a new energy scale labelled \textit{intermediate state energy} $(E-E_V)$. The error bar on the absolute value of the energy was set to 0.1 eV, which is the error on the evaluation of the zero-energy of the reference system.





The time evolution of the A, B, and C populations, shown in Fig. 3 in the main text, was fitted by a sum of exponential decay that considers the rise time and the decay times of the signal.
The response function is then convoluted with a Gaussian function, accounting for the time resolution of 200 fs.
To globally fit  the A, B, and C intensity, the decay of A is kept equal to the rise of B, and the two decays of B are equal to the two rise times of the C resonance.
The time resolution was corroborated by fitting the CC feature.


The exciton wave function,  and the total PE intensity are connected as follows. 
The total PE intensity is proportional to the squared transition dipole matrix element $|M^k_{f,i}|^2=|<\psi_f|\mathbf{A}\cdot \mathbf{p}|\psi_i>|^2$, which connects the initial state to the final state via the polarization operator $\mathbf{A}\cdot \mathbf{p}$, where $\mathbf{A}$ is the vector potential of the light field and $\mathbf{p}$ is the momentum operator. 
Using the assumption that the final state is a plane wave (plane wave approximation), the squared matrix element becomes $|M^k_{f,i}|^2\propto|\mathbf{A}\cdot \mathbf{k}|^2|<e^{-i\mathbf{k}\mathbf{r}}|\psi_i>|^2$. This expression indicates that the matrix element is proportional to the amplitude of the Fourier transform (FT) of the initial state wave function. \cite{Dong2021}.
The employed procedure used to retrieve the real space probability density of the electron contribution to the two-particle exciton wave function  and the resulting map is illustrated in Fig.5(d) in the main text.

\section*{Theoretical calculations}

\subsection*{Integrated band structure along $k_{z}$ direction of bulk BiI$_{3}$}

\begin{figure*}[h!]
\includegraphics[width=0.75\linewidth]{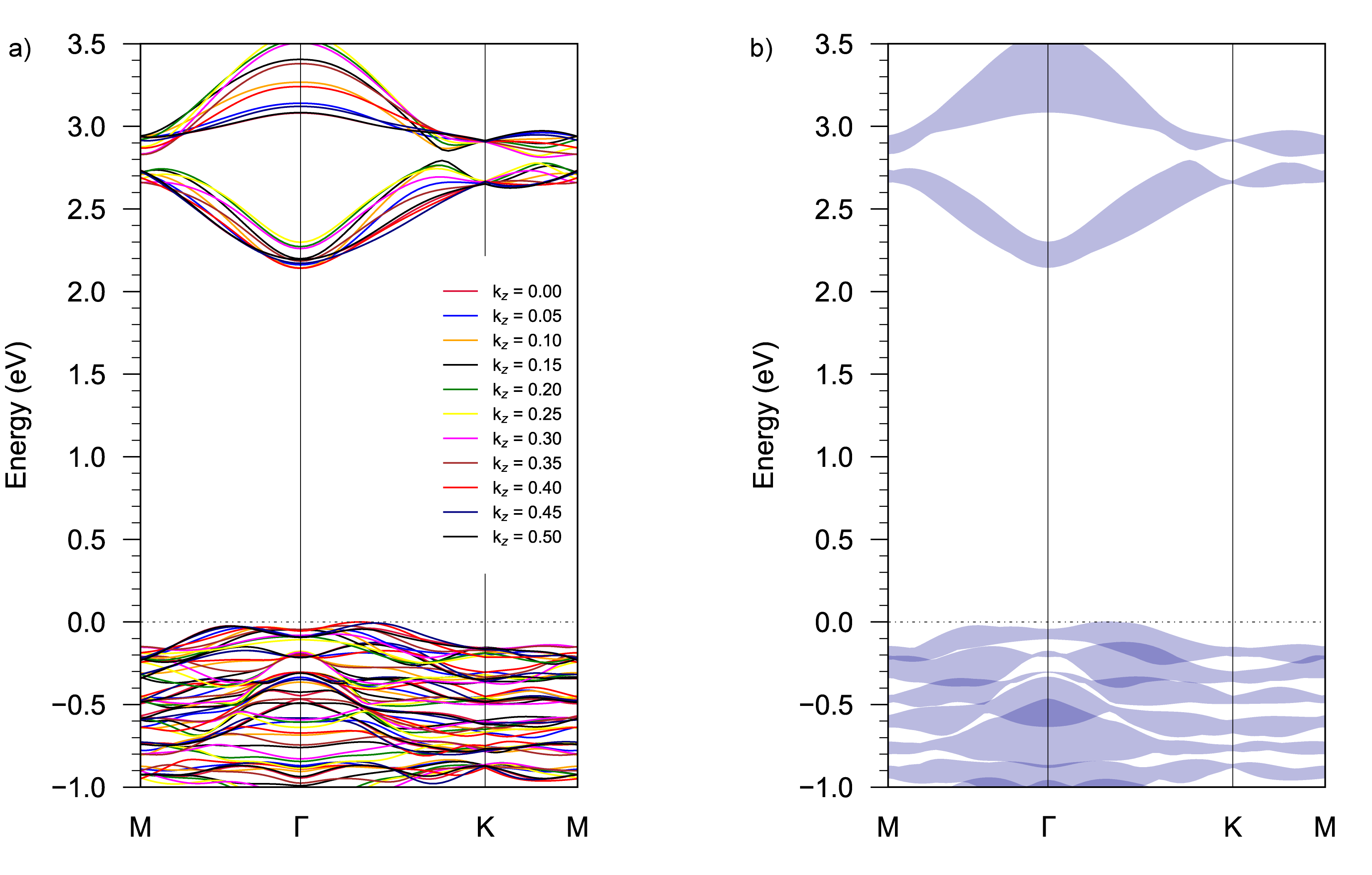}
\caption{\label{Surface-projected_BS} Calculation of the surface-projected band structure of bulk BiI$_{3}$. (a) Different band structures for each value of k$_{z}$. The values of k$_{z}$ are in crystal coordinates, where k$_{z} = 0.50$ is the first value in the second Brillouin zone. (b) Surface-projected band structure obtained from the integration of different band structures along the k$_{z}$ direction.}
\end{figure*}

Experimental measurements of tr-ARPES of bulk BiI$_{3}$ are performed over a parallel momentum region around $\Gamma$. To compare theoretical and experimental results, the surface-projected band structure is needed. To this end, we first obtain the hexagonal-projected path in the rhombohedral Brillouin zone. Then, we compute the band structure along the same path for k$_{x}$ and k$_{y}$ points of the reciprocal space and we vary the value of k$_{z}$, as depicted in Fig. \ref{Surface-projected_BS}(a). In this way, the surface-projected band structure shown in Fig. \ref{Surface-projected_BS}(b) is obtained by integrating all the values along k$_{z}$ direction.

\subsection*{Theoretical reconstruction of the tr-ARPES signal}

The reconstruction of the tr-ARPES signal was carried out using a GW@DFT+BSE scheme as a starting point. The electronic properties of BiI$_{3}$ at the ground-state level were calculated in the framework of DFT, as implemented in the Quantum ESPRESSO package \cite{giannozzi2009quantum,giannozzi2017advanced,giannozzi2020quantum}. Quasiparticle corrections were computed through the GW approximation while the optical properties considering excitonic effects were calculated by solving the finite-momentum BSE, as implemented in the Yambo code \cite{marini_yambo_2009,sangalli2019many,marsili2021spinorial} and reported in Ref \cite{Cervantes-Villanueva2024}. To model the tr-ARPES signal, we employ the expression from Ref \cite{Sangalli2021} given by:
\begin{equation}\label{trARPES_theo}
I[N_{\lambda \textbf{q}}(t)](\textbf{k}, \omega) = 2 \pi \sum_{\lambda \textbf{q}}  N_{\lambda \textbf{q}}(t) \sum_{cv} |A^{\lambda \textbf{q}}_{c v \textbf{k}}|^{2} \delta \left(\omega - (\epsilon_{v \textbf{k} - \textbf{q}} + \omega_{\lambda \textbf{q}})\right)
\end{equation}
where $t$ is the time, \textbf{k} identifies the electron momentum in the Brillouin zone, $\omega$ the energy of the tr-ARPES signal, $\lambda$ the exciton index, \textbf{q} the exciton momentum, v and c are the valence and conduction bands indexes. $N_{\lambda \textbf{q}}$(t) describes the exciton population-distribution as a function of the time, $A^{\lambda \textbf{q}}_{c v \textbf{k}}$ is the the exciton eigen-fuction in transition space, $\epsilon_{n\mathbf{k}}$ the band energy. $\epsilon_{v \textbf{k} - \textbf{q}}$ is obtained from the GW calculation, while $A^{\lambda \textbf{q}}_{c v \textbf{k}}$ and $\omega_{\lambda \textbf{q}}$ are collected when solving the finite-momentum BSE. The time evolution is not ab-initio, but defined via the following time-dependent excitonic distribution:
\begin{equation}
N_{\lambda \textbf{q}} (t) =  G_{\lambda \textbf{q}}(t) \alpha(t) + [1 - \alpha(t)]B_{\lambda \textbf{q}}(t) \hspace{5mm}
\end{equation}
where $G_{\lambda \textbf{q}}(t)$ is a Gaussian function, $B_{\lambda \textbf{q}}(t)$ is a Boltzmann distribution function and $\alpha(t)$ defines the weight of each distribution along the time. The Gaussian distribution is defined as a wave-packet propagating in energy and spreading both in energy and in momentum space:
\begin{equation}\label{GaussDist}
G_{\lambda \textbf{q}}(t) = \frac{1}{2\pi \sigma_{\omega}(t) \sigma_{\textbf{q}}(t)} exp \biggl[ -\frac{1}{2} \biggl(\frac{\omega_{\lambda \textbf{q}} - \omega_{0}(t)}{\sigma_{\omega}(t)} \biggr)^2 \biggr] exp \biggl[ -\frac{1}{2} \biggl(\frac{|\textbf{q}|^{2} - |\textbf{q}_{\omega_{0}}|^{2}}{\sigma_{\textbf{q}}(t)} \biggr)^2 \biggr] .
\end{equation}
$G_{\lambda \textbf{q}}(t)$ is centered on $\omega_{0}(t)$, 
which defines the time dependent propagation of the wave-packet. It is defined as
\begin{equation}\label{PumpEnergy}
\omega_{0}(t) = \omega_{0}-(\omega_{0}-\omega_{min}) \frac{Xt}{T}
\end{equation}
with $w_{0}$ the initial ($t = 0$) energy, corresponding to the pump energy, $\omega_{min}$ the lowest energy exciton, and $X/T$ is the velocity of propagation in energy space.
This simulates a time-linear excitonic propagation towards the lowest energy exciton.
$|\textbf{q}_{\omega_{0}}|^{2}\approx 0$ is the momentum at which the distribution is centered. $\sigma_{\omega}(t)$ and $\sigma_{\textbf{q}}(t)$ define the time-dependent energy and momentum broadening. The energy and momentum broadening, $\sigma_{\omega}(t)$ and $\sigma_{\textbf{q}}(t)$, are defined as: 
\begin{equation}
\sigma_{\omega}(t) = \frac{\sqrt{t}}{b} + \sigma_{\omega,0} \hspace{10mm} \sigma_{\textbf{q}}(t) = \frac{\sqrt{t}}{c} + \sigma_{\textbf{q},0}
\end{equation}
with $\sigma_{\omega,0}$ and $\sigma_{\textbf{q},0}$ the initial values. The variables $b$ and $c$ control a spreading proportional to the time square root.

On the other hand, the Boltzmann distribution, $B_{\lambda \textbf{q}}(t)$, is defined such that:
\begin{eqnarray}
B_{\lambda \textbf{q}}(t) = \frac{1}{Z}\, exp \biggl[ -\frac{\omega_{\lambda \mathbf{q}}}{k_B T(t)} \biggr] \\
Z(t) = \sum_{\lambda'\mathbf{q}'} exp \biggl[- \frac{\omega_{\lambda'\mathbf{q}'}}{k_B T(t)} \biggr]
\end{eqnarray}
where $k_B$ is the Boltzmann constant, $\omega_{\lambda\mathbf{q}}$ the exciton anergy and $T(t)$ is the time-dependent temperature. The dependency of the time for the temperature is given by:
\begin{equation}
T(t) = T_0 \cdot exp[-t \cdot f]
\end{equation}
where $T_0$ defines the initial value of the temperature at $t = 0$ and $f$ the decay ratio of the temperature with the time evolution. 

Finally, the function $\alpha(t)$ which determines the weight of each distribution is defined as:
\begin{equation}
\alpha(t) = e^{-t/\tau}
\end{equation}
defined in such a way that we have a purely gaussian wave-packet at the initial time, e.g. $\alpha(0)=1$ and a purely Boltzmann distribution at longer times, e.g. $\alpha(t)\approx 0$ for $t\gg\tau$.

\begin{figure*}[h!]
\includegraphics[width=0.90\linewidth]{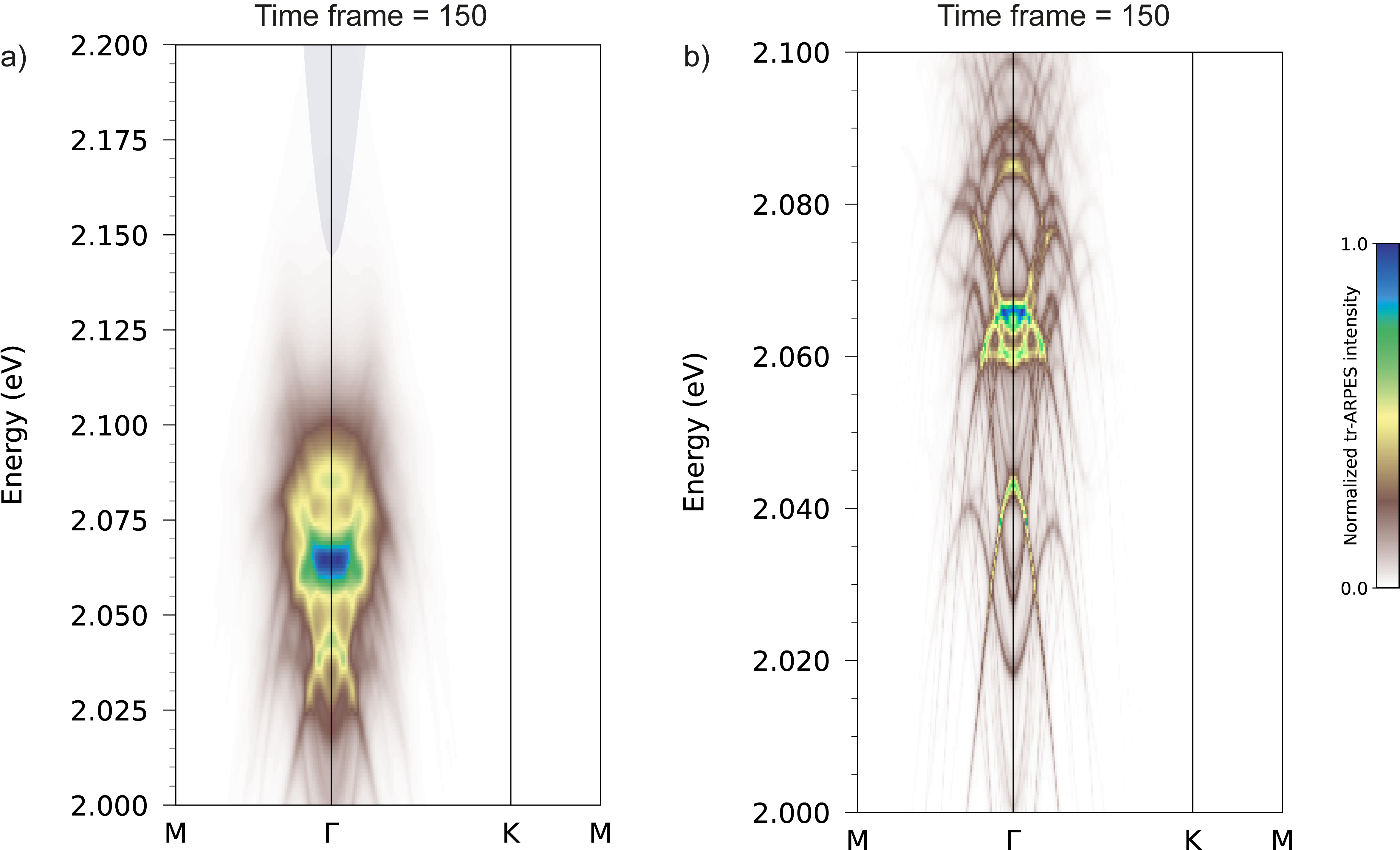}
\caption{\label{Bresonance_theo} Theoretical tr-ARPES signal associated with B resonance obtained using a smearing of (a) 0.005 eV, (b) 0.001 eV.}
\end{figure*}

\subsection*{Analysis of B resonance formation}

To understand more comprehensively the formation and shape of the B resonance observed experimentally, we analyse the contributions that yield the theoretical B resonance. For this purpose, we compute the tr-ARPES signal as in Fig. 4(c) of the manuscript (time frame = 150 and $T(t) = 139$) using smaller smearings, so that we can infer which excitonic and band states originate the B resonance. The results are depicted in \Fref{Bresonance_theo}, where from \Fref{Bresonance_theo}(a) we infer that the signal is more likely to present a concavity switch with respect to A resonance, in agreement with the experiment. A signal with smoother smearing is shown in \Fref{Bresonance_theo}(b), showing that although the concavity switch is still present, it has a very complex origin. Indeed, the B resonance is composed of many excitonic and valence bands, unlike in transition metal dichalcogenides, where the signal presents a dominant contribution \cite{PhysRevResearch.4.043203,liu2023time}. This is due to the small energy difference between the valence and excitonic bands, revealing many different non-zero contributions. As a result, the experimental effective mass for the B resonance cannot be assigned to a specific excitonic or valence band.